\documentclass[times,twocolumn,final]{elsarticle}

\usepackage{hyperref}
\usepackage{amssymb}
\usepackage{latexsym}
\usepackage{url}
\usepackage{xcolor}
\usepackage{pifont}
\usepackage{array}
\usepackage[acronym]{glossaries}

\usepackage{cleveref}

\crefname{figure}{Fig.}{Figs.}
\Crefname{figure}{Figure}{Figures}
	
\definecolor{newcolor}{rgb}{.8,.349,.1}

\newacronym{ai}{AI}{Artificial Intelligence}
\newacronym{midog}{MIDOG}{MItosis DOmain Generalization}
\newacronym{miccai}{MICCAI}{International Conference on Medical Image Computing and Computer Assisted Intervention}
\newacronym{wsi}{WSI}{Whole Slide Image}
\newacronym{mc}{MC}{mitotic count}
\newacronym{nmf}{NMF}{non-mitotic figures}
\newacronym{fov}{FOV}{field of view}
\newacronym{he}{H\&E}{hematoxylin and eosin}
\newacronym{gan}{GAN}{generative adversarial network}
\newacronym{roi}{ROI}{regions of interest}
\newcommand{\cmark}{\ding{51}}%
\newcommand{\xmark}{\ding{55}}%
\newcommand\snm[1]{#1}
\newcommand{\etal}{\textit{et~al.\-}}
\newacronym{CNN}{CNN}{convolutional neural network}
\newacronym{auprc}{AUPRC}{area under the precision recall curve}

\begin{document}
\begin{frontmatter}
\title{Mitosis domain generalization in histopathology images - The MIDOG challenge}
\author[1]{Marc \snm{Aubreville}\corref{cor1}}
\cortext[cor1]{Corresponding author: 
  Email: marc.aubreville@thi.de;}
\cortext[cor2]{Authors contributed equally.}
\author[2]{Nikolas \snm{Stathonikos}}
\author[3]{Christof A. \snm{Bertram}}
\author[4]{Robert \snm{Klopfleisch}}
\author[2]{Natalie \snm{ter Hoeve}}
\author[5]{Francesco \snm{Ciompi}}
\author[6]{Frauke \snm{Wilm}}
\author[6]{Christian \snm{Marzahl}}
\author[9]{Taryn A. \snm{Donovan}}
\author[6]{Andreas \snm{Maier}}
\author[p1]{Jack \snm{Breen}}
\author[p1]{Nishant \snm{Ravikumar}}
\author[p2]{Youjin \snm{Chung}}
\author[p2]{Jinah \snm{Park}}
\author[p3a]{Ramin \snm{Nateghi}}
\author[p3b]{Fattaneh \snm{Pourakpour}}
\author[p4]{Rutger H.J. \snm{Fick}}
\author[p4]{Saima \snm{Ben Hadj}}
\author[p5]{Mostafa \snm{Jahanifar}}
\author[p5]{Nasir \snm{Rajpoot}}
\author[p6]{Jakob \snm{Dexl}}
\author[p6]{Thomas \snm{Wittenberg}}
\author[p7]{Satoshi \snm{Kondo}}
\author[p8]{Maxime W. \snm{Lafarge}}
\author[p8]{Viktor H. \snm{Koelzer}}
\author[p9]{Jingtang \snm{Liang}}
\author[p9]{Yubo \snm{Wang}}
\author[p10]{Xi \snm{Long}}
\author[p10b]{Jingxin \snm{Liu}}
\author[p12]{Salar \snm{Razavi}}
\author[p12]{April \snm{Khademi}}
\author[p13a]{Sen \snm{Yang}}
\author[p13b]{Xiyue \snm{Wang}}

\author[7]{Mitko \snm{Veta}\corref{cor2}}
\author[8]{Katharina \snm{Breininger}\corref{cor2}}
\address[1]{Technische Hochschule Ingolstadt, Ingolstadt, Germany}
\address[2]{Pathology Department, UMC Utrecht, The Netherlands}
\address[3]{Institute of Pathology, University of Veterinary Medicine, Vienna, Austria}
\address[4]{Institute of Veterinary Pathology, Freie Universit{\"a}t Berlin, Berlin, Germany}
\address[5]{Computational Pathology Group, Radboud UMC Nijmegen, The Netherlands}
\address[6]{Pattern Recognition Lab, Friedrich-Alexander-Universität Erlangen-Nürnberg, Germany}
\address[9]{Department of Anatomic Pathology, Schwarzman Animal Medical Center, New York, USA}
\address[p1]{CISTIB Center for Computational Imaging and Simulation Technologies in Biomedicine, School of Computing, University of Leeds, UK}
\address[p2]{Korea Advanced Institute of Science and Technology, Daejeon, South Korea}
\address[p3a]{Electrical and Electronics Engineering Department, Shiraz University of Technology, Shiraz, Iran}
\address[p3b]{Iranian Brain Mapping Biobank (IBMB), National Brain Mapping Laboratory (NBML), Tehran, Iran}
\address[p4]{Tribun Health, Paris, France}
\address[p5]{Tissue Image Analytics Centre, Department of Computer Science, University of Warwick, UK}
\address[p6]{Fraunhofer-Institute for Integrated Circuits IIS: Erlangen, Germany}
\address[p7]{Muroran Institute of Technology, Hokkaido, Japan}
\address[p8]{Department of Pathology and Molecular Pathology, University Hospital and University of Zurich, Zurich, Switzerland}
\address[p9]{School of Life Science and Technology, Xidian University, Shannxi, China}
\address[p10]{Histo Pathology Diagnostic Center, Shanghai, China}
\address[p10b]{Xi'an Jiaotong-Liverpool University, Suzhou, China}
\address[p12]{Image Analysis in Medicine Lab (IAMLAB), Electrical, Computer and Biomedical Engineering, Ryerson University, Toronto, ON, Canada}
\address[p13a]{Tencent AI Lab, Shenzhen 518057, China}
\address[p13b]{College of Computer Science, Sichuan University, Chengdu 610065, China}

\address[7]{Medical Image Analysis Group, TU Eindhoven, The Netherlands}
\address[8]{Department of Artificial Intelligence in Biomedical Engineering, Friedrich-Alexander-Universität Erlangen-Nürnberg, Germany}
\date{March 2022}

\begin{abstract}
The density of mitotic figures within tumor tissue is known to be highly correlated with tumor proliferation and thus is an important marker in tumor grading. Recognition of mitotic figures by pathologists is known to be subject to a strong inter-rater bias, which limits the prognostic value. State-of-the-art deep learning methods can support the expert in this assessment but are known to strongly deteriorate when applied in a different clinical environment than was used for training. One decisive component in the underlying domain shift has been identified as the variability caused by using different whole slide scanners. The goal of the MICCAI MIDOG 2021 challenge has been to propose and evaluate methods that counter this domain shift and derive scanner-agnostic mitosis detection algorithms. The challenge used a training set of 200 cases, split across four scanning systems. As a test set, an additional 100 cases split across four scanning systems, including two previously unseen scanners, were given. The best approaches performed on an expert level, with the winning algorithm yielding an $F_1$ score of 0.748 (CI95: 0.704-0.781). In this paper, we evaluate and compare the approaches that were submitted to the challenge and identify methodological factors contributing to better performance.
\end{abstract}

\end{frontmatter}

\section{Introduction}
Deep learning has revolutionized the field of digital histopathology in recent years, as methods continue to emerge that perform on par or even surpass the performance of human experts in specific tasks \citep{levine2019rise,karimi2019deep,aubreville2020deep}. The application of  \gls{ai} methods in a computer-aided diagnostic workflow is especially beneficial for quantitative routine tasks, allowing for a faster diagnostic process, or for tasks with a known high inter-rater variability to increase diagnostic reproducibility by reducing diagnostic bias. 

One task for which both of these conditions are met is tumor grading, i.e. the assessment of the malignant potential of a tumor from histological specimens \citep{veta2015assessment,balkenhol2019deep}. Many tumor grading schemes rely on the identification and the counting of cells in the process of cell division (mitotic figures). The density of mitotic figures in an area (mitotic activity) is known to be highly correlated with proliferation of the tumor \citep{Baak:2008cma}. Yet, identification of mitotic figures is known to suffer from a significant inter-rater variability \citep{Meyer:2005cl,Meyer:2009eu}, which might be the dominant limiting factor for the prognostic value. Studies have shown that by using \gls{ai}-based methods both reproducibility and accuracy can be increased \citep{Bertram2021VetPathol,balkenhol2019deep}. 

One major limitation of the state-of-the-art deep learning-based methods is that their performance is known to significantly deteriorate with a covariate shift of the images, i.e. a change in visual representation between images that the model was trained upon and those that it encounters during inference in a clinical diagnostic workflow. Contrary to machine learning models, humans can often adapt seamlessly to this shift \citep{stacke2020measuring,aubreville2021quantifying}. The main causes for such a domain shift in histopathology are the staining procedure (which can differ over time and/or across laboratories), the acquisition device (whole slide scanner), and the tumor type itself (different tumor cell morphology and tissue architecture). While a limitation of an algorithmic aid to a specific subclass of tissues may be an acceptable restriction, a limitation towards factors that characterize a laboratory environment (tissue preparation, staining procedure, scanner) prevents the use of such methods in diagnostic practice across labs. 

\begin{figure}
    \centering
    \includegraphics[width=0.48\textwidth]{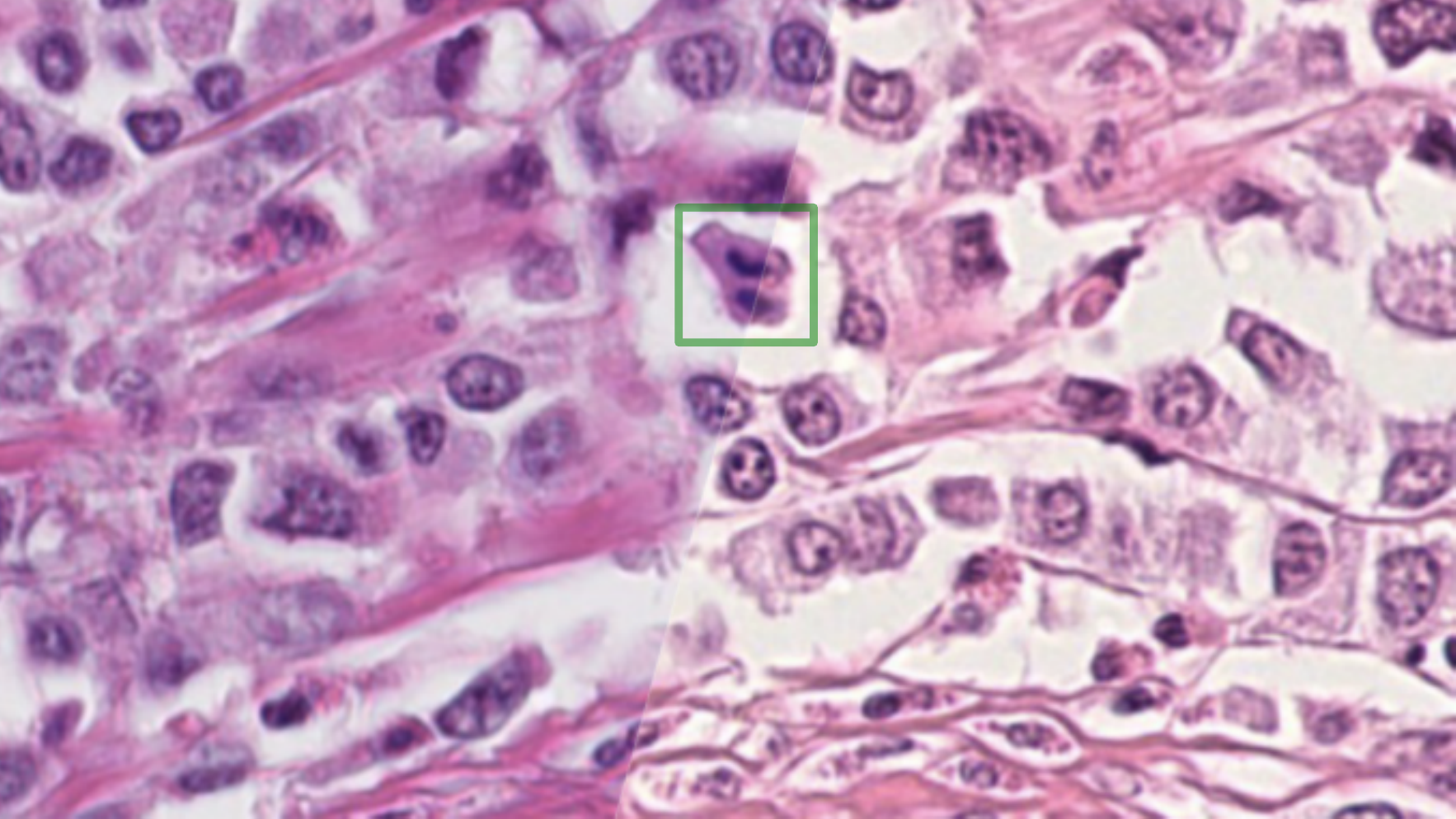}
    \caption{Breast cancer tissue, acquired using a Hamamatsu NanoZoomer XR (left, Scanner A) and a Hamamatsu NanoZoomer S360 (right, Scanner B). Besides a color shift, the depth of field is also affected by using a different scanner, caused by different optical properties of the objective. A mitotic figure in anaphase is indicated by a green box.}
    \label{fig:example_scanners}
\end{figure}

This motivated the design of the \gls{midog} 2021 challenge. As the \gls{wsi} scanner used for digitization was identified to cause a strong domain shift \citep{aubreville2021quantifying}, presumably even stronger than the domain shift caused by the tissue preparation and staining procedure \citep{aubreville2020completely}, the challenge focused on the task of generalizing against scanner-induced domain shift for the identification of mitotic figures in histopathology images. Detection of mitotic figures is prone to shifts of color representation of the digital image, since both color and structural patterns are required for a proper identification. Besides color representation, which is influenced by the light source and sensors of the \gls{wsi} scanner and proprietary color calibration schemes, the optical parameters of the microscope, such as the numerical aperture, also influence the representation of mitotic figures in the digital image (see  \cref{fig:example_scanners}).

\begin{figure*}
\centering
\includegraphics[width=\textwidth]{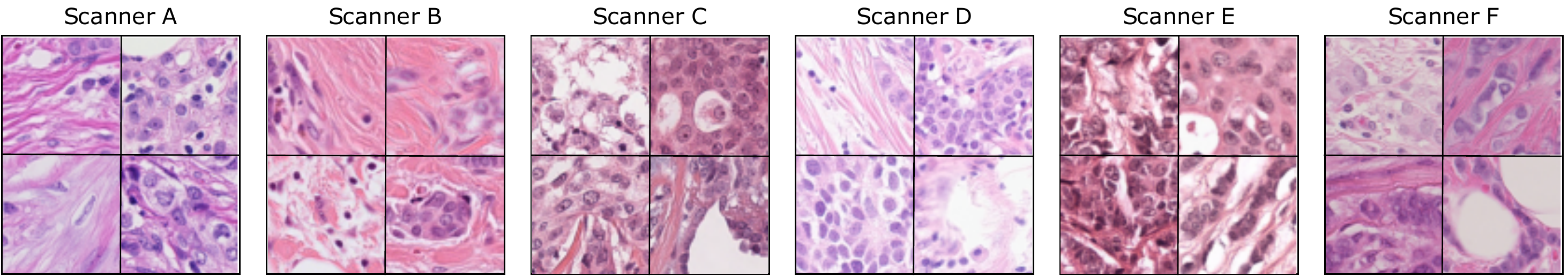}
\caption{Random crops of breast cancer tissue from all six scanners of the training and test set, showing the visual variability between the scanners. }
\label{fig:scanners}
\end{figure*}

\subsection*{Challenge format and task}
The challenge was held in conjunction with the 2021 \gls{miccai} conference as a one-time event. The structured challenge description \citep{challengedescription} was accepted after a single-blind peer review process. Participants were able to register and obtain the training data five months prior to the challenge submission deadline, allowing for sufficient time to develop and evaluate their algorithms. After registration, the participants were provided with the training set data and a description of the images and annotations, together with a Jupyter notebook showcasing how to work with the data and how to train an example object detector (RetinaNet) with it. 

The task of the challenge was the detection of mitotic figures on \glspl{roi} of a predefined size. For the training set, 200 cases of breast cancer (representing 200 patients) were retrieved, the \gls{roi} was selected by a trained pathologist and the \glspl{wsi} were digitized using four different \gls{wsi} scanners (50 cases each). The test set consisted of 80 cases digitized by four scanners (20 each) out of which two scanners were part of the training set and two were unseen. The challenge dataset represents a good trade-off between capturing the naturally occurring variability of \glspl{wsi} and time invested for annotation \citep{challengedescription}. The number of cases per scanner allows for a realistic estimation of the performance. Due to the task being about generalization, prior knowledge about the images of the test set needed to be excluded. Thus, the participants had no knowledge of the nature of the test scanners, no access to the test images and had to submit a Docker container to be evaluated automatically on the test data. For this, a reference algorithm \citep{wilm2022} embedded into a Docker container\footnote{\url{https://github.com/DeepPathology/MIDOG_reference_docker}} was made available to the participants alongside a textual description and video tutorials on how to use it.

All participating teams submitted their fully automatic algorithm containers on the grand-challenge.org website. To check container functionality and algorithmic validity, a preliminary test set (20 cases, 5 per scanner, same scanners as in the test set) was made available for automatic evaluation on the platform, two weeks prior to the submission deadline. The evaluation container for the challenge was also made available on github\footnote{\url{https://github.com/DeepPathology/MIDOG_evaluation_docker/}}. Participants were not permitted to use other sources of images to train their models, besides general purpose datasets such as ImageNet. In order to prevent overfitting to the characteristics of the preliminary test seft, the number of submissions was limited to one per day. After 15 days, the preliminary test phase was closed and the final test phase was started, to which the participants were able to only submit once. Alongside with the submission for evaluation on the final test set, participating teams had to provide a short paper, describing their approach and their preliminary results, on a publicly available pre-print server. Further details about the challenge, including details about the submission instructions, the publication policy and the timetable can be found in the structured challenge description \citep{challengedescription}.

Initially, 237 individuals registered on the challenge website\-\footnote{\url{https://imig.science/midog/}} and 161 users joined the challenge on the grand-challenge.org platform. Members of the organizers' institutes were not allowed to parcitipate in the challenge. 46 users submitted at least one docker container to the challenge. At the end, 17 teams made a submission to the final test set. Single-blind peer review was carried out on all submitted short papers. After the peer review, we invited all teams that passed peer review and where the approach exceeded a minimum score of $F_1\geq0.6$ to the workshop (12/17 teams, acceptance rate 70.6\%). The approaches presented in the workshop also are compared in this paper.

\section{Material and methods}

\noindent The main design principles of the challenge were:
\begin{enumerate}
    \item To have a generally representative dataset of a relevant disease and an important diagnostic task.
    \item To reflect a real clinical use case by ensuring a truly independent hold out set with unknown characteristics.
    \item To achieve high label quality to ensure an accurate evaluation.
\end{enumerate}

\noindent 
With around 2.3 million cases in 2020, breast cancer is one of the cancer types with the highest prevalence \citep{wild2020world}. Patients can benefit from adjuvant therapies significantly, However, aggressive therapies also carry the risk of serious side effects and thus should be restricted to patients with unfavorable prognostic markers, such as high tumors proliferation \citep{van2004prognostic}. One marker strongly correlated with proliferation is the mitotic count, i.e., the assessment of cells undergoing cell division (mitosis) in a defined area (commonly 10 high-power fields, here defined as $2.0mm^2$) \citep{Elston:1991dl}. This \gls{roi} is selected by an expert pathologist within the tumor area with the presumed highest mitotic activity in a hematoxylin and eosin-stained digitized microscopy slide. Since this task is part of many tumor grading schemes (e.g., meningioma \citep{Louis2016} or lung adenocarcinoma \citep{moreira2020grading}), it is highly relevant for general tumor prognostication and was chosen as target task for our challenge.

In order to ensure representativeness of the dataset, we chose a rigorous inclusion scheme for the challenge cohort. 
For the challenge to yield trustworthy results and especially to avoid a methodological overfitting, the independent test set needed to stay a true hold out, i.e. completely hidden to the participants. 

To ensure the quality of our evaluation, the labeling quality was improved by using multiple experts and, additionally, a machine-learning-augmented annotation pipeline (see below).

\subsection{Challenge cohort}
The challenge dataset consists of 300 breast cancer cases and was curated from a retrospective, consecutive selection taken from the diagnostic archive of the University Medical Center (UMC) Utrecht, The Netherlands. All samples were resected  solely for diagnostic purposes.  Inclusion criteria were the availability of the microscopy (glass) slide, a confirmed breast cancer excision (as documented in the patient record), the availability of a pathology report with a documented mitotic count and that the patients did not opt out for the use of their data in research projects. Prior to handing over to project partners within the organization committee, all slides and clinical meta data were anonymized. Additionally, for the use in the challenge we obtained approval by the institutional review board of UMC Utrecht (reference: TCBio 20-776). The specimen was preprocessed according to clinical standard routine and stained with \gls{he} dye. Subsequently, the cases were randomly split into the training set (200 cases) and the preliminary (20 cases) and final (80 cases)
test set. Within each of those sets, we performed another random split, to assign the cases to the scanners, i.e. we split up the 200 cases of the training set into 50 cases for each of the four training scanners (A,B,C,D, see below), and the 80+20 cases of the test sets into 20+5 for each of the the four test scanners (A,D,E,F, see below).  By this procedure, we can expect no significant biases in any of the subsets of the dataset and assume a high degree of representativeness. 

\subsection{Image acquisition}
We used four different scanners for the digitization of the training set:
\begin{itemize}
    \item Scanner A: Hamamatsu NanoZoomer XR (C12000-22, Hamamatsu, Hamamatsu City, Japan), optical resolution: 0.23 microns/px at 40x magnification
    \item Scanner B: Hamamatsu NanoZoomer S360 (Hamamatsu, Hamamatsu City, Japan), optical resolution: 0.23 microns/px at 40x magnification
    \item Scanner C: Aperio Scanscope CS2 (Leica Biosystems, Nussloch, Germany), optical resolution: 0.25 microns/px at 40x magnification
    \item Scanner D: Leica Aperio GT 450 (Leica Biosystems, Nussloch, Germany), optical resolution: 0.26 microns/px at 40x magnification, custom optics by Leica Microsystems for native 40x scanning with 1 mm \gls{fov}
\end{itemize}

Scanner A is the scanner that is used in clinical practice to digitize all slides at UMC Utrecht. Therefore, all slides of our dataset are also available scanned by this scanner and we use this as our reference scanner to counter a possible scanner-caused bias in the region of interest selection.

For the test set, we used a mix of known and unknown scanners to test simultaneously for performance on in-domain scanners and for generalization to out-of-domain scanners. In addition to the scanners A and D, the test set was scanned with:
\begin{itemize}
    \item Scanner E: 3DHISTECH Panoramic 1000 (3DHISTECH, Budapest, Hungary), optical resolution: 0.24 microns/px at 20x magnification, Plan-Apochromat objective, numerical aperture of lens: 0.8
    \item Scanner F: Hamamatsu NanoZoomer 2.0RS (C10730-12, Hamamatsu, Hamamatsu City, Japan), optical resolution: 0.23 microns/px at 40x magnification, numerical aperture of lens: 0.75
\end{itemize}

For scanner D, which was part of the training and the test set, no labels were provided as part of the training set. Hence, this scanner was included for the sole purpose of providing data for approaches performing unsupervised domain adaptation. 

Each of the slides was scanned by the assigned \gls{wsi} scanner as well as by the clinical reference scanner (Scanner A). To reduce bias which might be caused by different tissue representation on other scanners, a trained pathologist (C.B.) selected a region of interest spanning $2.0\,mm^2$ on the reference scans. For streamlined dataset creation, all \glspl{wsi} were uploaded to a central server running a collaborative annotation software \citep{marzahl2021exact}. There, the reference scans were registered to the respective image acquired by different scanners using a quadtree-based \gls{wsi} registration method by Marzahl \etal \citep{marzahl2021robust}. The registration was manually fine-tuned and quality checked subsequently. Finally, the defined \glspl{roi} were extracted from the images acquired by all scanners. 

\begin{figure}
    \centering
    \includegraphics[width=0.48\textwidth]{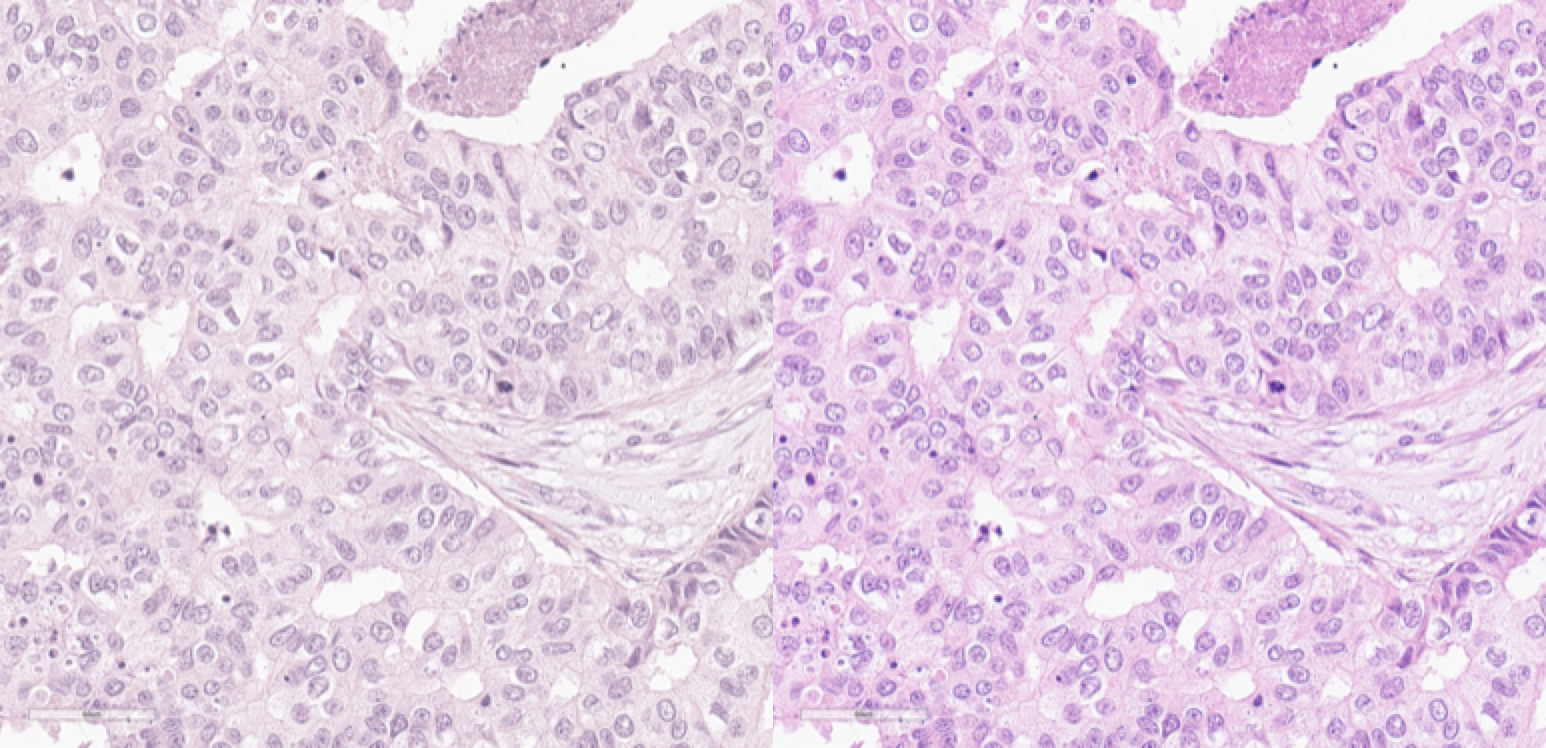}
    \caption{Image from Scanner D in the original (left) scanner color space and in the standard RGB color space (right).}
    \label{fig:scanner_calib}
\end{figure}

For the Leica and Aperio scanner, the scanners color profile was available from the \glspl{wsi}, which enabled the organizers to convert those images into the standard RGB color space. Regardless of this step, significant differences in color can be observed (see \cref{fig:scanner_calib}).

\begin{figure*}[!t]
    \centering
    \includegraphics[width=\textwidth]{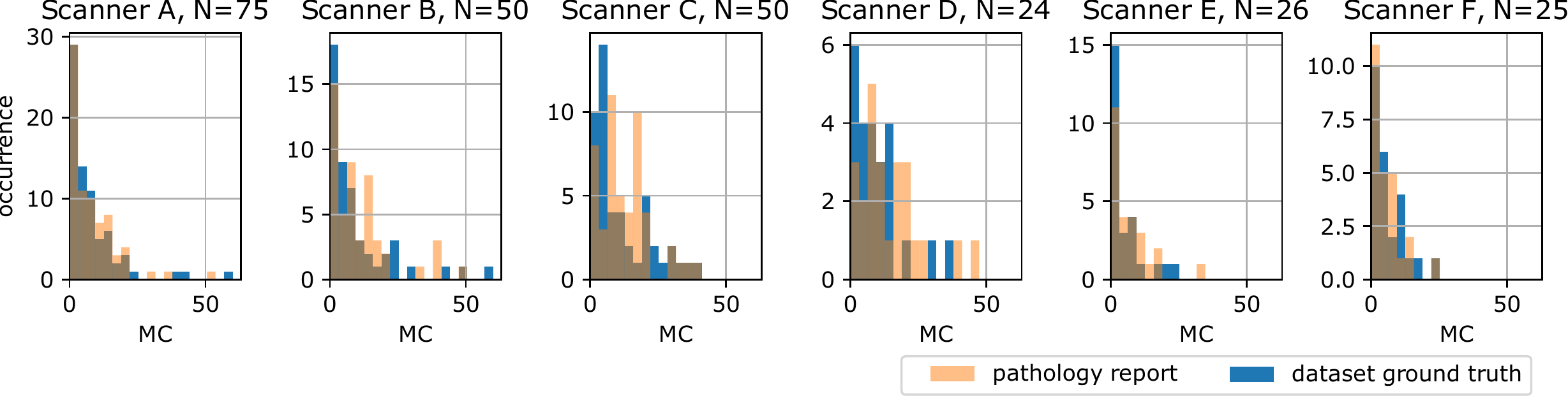}
    \caption{Distribution (histogram) of mitotic count (MC) across cases and scanners, according to the the pathology report (orange) and the MIDOG dataset (blue).}
    \label{fig:histo_dataset}
\end{figure*}

\subsection{Labeling}

In the process of labeling, three expert pathologists (C.B., R.K., T.D.) were involved. All experts work at different institutions (in different countries) and have 6+ years of professional experience and a demonstrably high level of expertise in mitotic figure recognition. Additionally, all experts agreed on common criteria for mitotic figure identification \citep{donovan2021mitotic}.

The inter-rater variability of mitosis identification can be attributed to two main factors: Most notably, experts disagree on individual mitotic figures (object-level disagreement) due to morphological overlap with imposters \citep{veta2016mitosis}. For this reason, most mitotic figures datasets were annotated as consensus voting between multiple experts \citep{veta2015assessment,veta2019predicting,roux2014mitos,bertram2019large,bertram2020pathologist,aubreville2020completely}. Additionally, as mitotic figures can be considered sparse events in most \glspl{wsi} and sometimes faint structures, experts tend to miss especially less recognizable objects when screening the image \citep{Bertram2021VetPathol,bertram2019large}. 

To account for both effects and create a high-quality dataset, we employed a machine learning-aided strategy \citep{bertram2020pathologist}: Initially, a single experienced expert (C.B.) screened all \glspl{roi} for mitotic figures and for a roughly equal number of imposters / hard negatives (non-mitotic cells with morphological similarity to mitotic figures). We trained a customized \citep{marzahl2020deep} RetinaNet \citep{lin2017focal} to identify cells that were missed in this initial labeling (mitotic figure candidates). Using a very low detection threshold on the model output, we ensured that this approach yielded high sensitivity (and low specificity). All mitotic figure candidates were then handed to the primary expert, again, to sort into missed mitotic figures and non-mitotic structures (which comprised the vast majority). 

To reduce bias in the labeling process introduced by the first expert, all manual cell labels and detections from the algorithmic augmentation step were class-blinded and handed to a secondary expert (R.K.). The secondary expert then sorted all cells into mitotic figures and non-mitotic cells. In the case of agreement, the label was accepted as ground truth. In case of disagreement, the cells were again class-blinded and given to a third expert (T.D.), who then made the final decision. 

In the dataset, the challenge organizers provided squared approximated bounding boxes of equal size (50px) to the participants. The non-mitotic cells (hard negatives) were provided alongside the true mitotic figures to enable the use within sampling schemes.

\subsection{Dataset statistics}
As \cref{fig:histo_dataset} shows, the \gls{mc} (i.e., the number of mitoses per $2.0\,mm^2$) follows a similar distribution over all scanners in the original pathology report and the MIDOG dataset. Differences between the \gls{mc} of the pathology report and the MIDOG dataset were not significant (two-sided paired t-test: p-value: 1.000, scipy stats package version 1.7.3). The intraclass correlation (ICC, two-way, single measurements, random raters) of the \gls{mc} ratings indicate a substantial agreement (ICC2=$0.684$, pingouin package version 0.5.1) as of the definition by Hallgren \citep{hallgren2012computing}, which can be attributed to different reasons. Firstly, it cannot be guaranteed that the slides that were included in this dataset were the same that were used for clinical \gls{mc} assessment, as oftentimes multiple slides per case exist. Secondly, the annotation methodology used here differs substantially from how the \gls{mc} is determined in a clinical setting. Lastly, even if the same slides were used, the \gls{mc} is known to be highly area-dependent, which might further contribute to the discordance \citep{bertram2020computerized}.   

As depicted in \cref{fig:histo_dataset}, there is a high number of potentially low-grade tumors within the dataset, which is, however, reflective of the general population of cases at a tertiary hospital. 

\subsection{Reference approach}
As a starting point, and to familiarize the participants with the submission process, the organizers provided a repository including an exemplary Docker container to all participants. The container provided a trained model including model weights and all scripts to run inference with it in the test environment. Alongside, the organizers made a description of the reference approach available to all participants during the challenge. The approach \citep{wilm2022} was based upon a customized \citep{marzahl2020deep} RetinaNet \citep{lin2017focal} implementation, where object classification and bounding box regression are solely performed at the highest resolution of the feature pyramid network. At the end of the encoder, a domain classification head was attached to the network. The task of the domain classification head is the discrimination of the four scanners of the training set. In between both, a gradient reversal layer \citep{ganin2016domain} was attached. This layer acts as a unity transform in the forward pass of the network, but inverts the networks' gradients, weighted by a constant, in the backward pass. This has the effect of adversarial training, i.e. of making the features less discriminative for the domain, and effectively reducing the domain covariate shift in feature space \citep{lafarge2019learning}. The model uses a combined loss with terms for domain discrimination, object classification, and bounding box regression, using focal loss \citep{lin2017focal} for both classification tasks and a smooth L1 loss for the regression task. For scanner D (i.e., the scanner without annotations), only the domain-adversarial part of the loss is active. Besides this domain adaptation technique, the model only uses standard image augmentation techniques (affine transforms, flipping, contrast and brightness adjustments). For model selection, the \gls{auprc} was calculated for a validation set, consisting of a selection of ten cases of the scanners from the training set where annotation data was available (scanners A,B, and C). The model with the highest average \gls{auprc} value was selected to be run on the test set. There was no access to any of the test sets during the development of the reference approach and no knowledge of the scanners selected for testing.

Additionally, \cite{wilm2022} provide a baseline just using standard augmentation, which we will refer to as CNN baseline in the following. This model is using the same network topology and training scheme as the Reference model, but missing domain-augmentation techniques besides standard image augmentation (brightness, contrast, random zoom, flipping and rotation).

\subsection{Evaluation methods}

\begin{table*}[!t]
\caption {\label{tab:methods} Overview of methods submitted to the MIDOG challenge.} 

\resizebox{\textwidth}{!}{
\begin{tabular}{p{20mm}>{\raggedright\arraybackslash}p{60mm}p{15mm}p{15mm}p{15mm}p{15mm}p{15mm}p{15mm}p{30mm}p{25mm}p{25mm}>{\raggedright\arraybackslash}p{20mm}>{\raggedright\arraybackslash}p{30mm}}

\hline
Team name & Core method/architecture & Multi-stage & Ensemble or TTA & \multicolumn{5}{c}{Data augmentation}    & \multicolumn{2}{c}{Domain adaptation} & Used unlabelled domain & Used additional labels \\ 
\cline{5-11}                                                          
 &  &  &  & Color & Staining  & Brightness & Contrast & Synthesis & Staining normalization & Other \\    \hline
 
Reference method \citep{wilm2022} & RetinaNet \cite{lin2017focal} with ResNet-18 \cite{he2016deep} encoder & \xmark & \xmark &  \xmark & \xmark & \cmark & \cmark & \xmark & \xmark & Domain-adversarial training \cite{pasqualino2021unsupervised} & \cmark & \xmark \\ 
CNN baseline \citep{wilm2022} & RetinaNet \cite{lin2017focal} with ResNet-18 \cite{he2016deep} encoder & \xmark & \xmark &  \xmark & \xmark & \cmark & \cmark & \xmark & \xmark & \xmark & \xmark & \xmark \\
\hline

AI medical \citep{midog_aimedical} & SK-UNet \cite{wang2021sk} & \xmark & \xmark & \cmark & \xmark & \cmark & \xmark & Fourier domain mixing \cite{yang2020fda} & \xmark & \xmark & \cmark & Mitosis segmentations \\

TIA Centre \citep{midog_tiacentre} & Efficient-UNet \cite{jahanifar2021robust} (candidate segmentation),  Efficient-Net-B7 \cite{tan2019efficientnet} (candidate classification) & \xmark & \cmark & \cmark & \xmark & \cmark & \cmark & \xmark & Vahadane et al. \cite{vahadane2016structure} & \xmark & \xmark & Mitosis segmentations \\

Tribun Healthcare \citep{midog_tribun} & Mask-RCNN \cite{he2017mask} (candidate detection), ResNet-50 \cite{he2016deep} and DenseNet-201 \cite{huang2017densely} (candidate classification) & \cmark & \cmark & \xmark & \xmark & \xmark & \xmark & CycleGAN \cite{de2021residual} & \xmark & \xmark & \cmark & Mitosis segmentations\\

CGV \citep{midog_cgv} & RetinaNet \cite{lin2017focal} with ResNet-101 \cite{he2016deep} encoder & \xmark & \xmark & \xmark & \xmark & \xmark & \xmark & StarGAN \cite{choi2018stargan} & \xmark & \xmark & \cmark & \xmark \\

XidianU-OUC \citep{midog_xidianu} & DetectorRS \cite{qiao2021detectors} (candidate detection), ensemble of 5 models for candidate classification & \cmark & \cmark & \xmark & \xmark & \cmark & \cmark & \xmark & Macenko et al. \cite{macenko2009method} & \xmark & \xmark & \xmark \\

IAMLAB \citep{midog_iamlab} & Cascade R-CNN \cite{cai2018cascade} with ResNet-101 \cite{he2016deep} encoder & \cmark & \xmark & \cmark & \xmark & \xmark & \cmark & \xmark & Macenko et al. \cite{macenko2009method} & \xmark & \xmark & \xmark \\

No.0 \citep{midog_no0} & Cascade R-CNN \cite{cai2018cascade} with ResNet-50 \cite{he2016deep} encoder & \cmark & \xmark & \xmark & \cmark & \xmark & \xmark & \xmark & \xmark & Domain-adversarial training with PatchGAN \cite{isola2017image} & \cmark & \xmark \\

jdex \citep{midog_jdex} & RetinaNet \cite{lin2017focal} with Efficient-Net-B0 \cite{tan2019efficientnet} encoder & \xmark & \xmark & \cmark & \cmark & \cmark & \cmark & \xmark & \xmark & \xmark & \xmark & \xmark \\

Leeds \citep{midog_leeds} & UNet with ResNet-152 \cite{he2016deep} encoder & \xmark & \xmark & \cmark & \xmark & \cmark & \cmark & \xmark & \xmark & \xmark & \xmark & \xmark \\

PixelPath-AI \citep{midog_pixelpathai} & Faster-RCNN \cite{ren2015faster} (candidate detection), Efficient-Net-B0 \cite{tan2019efficientnet} (candidate classification)  &\cmark & \cmark & \xmark & \cmark & \cmark & \cmark & \xmark & \xmark & \xmark & \xmark & \xmark \\

SK \citep{midog_sk} & Thresholding of the blue ratio image (candidate detection) \cite{chang2012nuclear}, ResNet \cite{he2016deep} (candidate classification) & \cmark & \xmark & \cmark & \xmark & \cmark & \cmark & \xmark & \xmark & Domain-adversarial training \cite{ganin2016domain} & \cmark & \xmark \\
ML \citep{midog_ml} & Rotation-invariant CNN \cite{lafarge2021roto} with 7 trainable layers & \xmark & \cmark & \cmark & \xmark & \cmark & \cmark & \xmark & \xmark & \xmark & \xmark & \xmark \\
\hline


\end{tabular}}
\end{table*}

All participants were required to submit their approach as a docker container to the grand-challenge.org platform. There, the containers were automatically evaluated for each image of the test set independently. The participants had no access to any of the test images during the challenge and detailed detection results were also not available to them. This was done in order to ensure a fair comparison of the approaches and to discourage overfitting or manual fine-tuning towards the test set. 

To ensure proper functionality of the automatically evaluated container images, we provided evaluation results on the preliminary test set. The main metrics ($F_1$ score, precision and recall) of submitted approaches were made available on a public leaderboard. 

For the overall rating, the $F_1$ score was the primary metric. We calculated the $F_1$ score over all $N$ processed slides as

$$ F_1 = \frac{2 \sum_k^N \mathrm{TP}_k}{2 \sum_k^N \mathrm{TP}_k + \sum_k^N \mathrm{FN}_k + \sum_k^N \mathrm{FP}_k}$$ 

where $\mathrm{TP}_k$, $\mathrm{FN}_k$, and $\mathrm{FP}_k$ are the number of true positive, false negative and false positive detections on slide $k$ and $N$ is the total number of slides ($N=80$ for the final test set and $N=20$ for the preliminary test set). 

The $F_1$ score was chosen as the main evaluation metric because it is defined as harmonic mean of precision and recall, and both underestimation as well as overestimation of the MC are equally severe for the diagnostic process. We did opt to calculate the overall $F_1$ instead of the mean $F_1$ of all slides, since slides with a low prevalence of mitotic figures would be overrepresented in this average.

We defined a detection to be a true positive whenever the Euclidean distance between a mitotic figure annotation and the detection was less than $7.5\,\mu m$. This value corresponds to the average size of mitotic figures in our dataset and provides a reasonable tolerance for misalignment of detection and ground truth labels. All detections not within $7.5\,\mu m$ of a ground truth mitotic figure annotation were considered false positives. Multiple detections of an already detected object were also counted as false positives, since they would introduce a positive bias to the \gls{mc}. All ground truth mitotic figures without a detection within a proximity of $7.5\,\mu m$ are considered false negatives. 

For the evaluation in this paper, we also calculated the $F_1$ metric for each scanner. Further, we performed bootstrapping process \citep{hall1994methodology} where we randomly selected M cases with replacement (where M is the number of images available per scanner or overall). This process was repeated 10,000 times to be able to derive a statistical distribution for the $F_1$ metric. This was also used to estimate the 5\% and 95\% confidence intervals for the $F_1$ score.

\subsection{Post-challenge ensembling of approaches}
We were also interested in discriminating parts of the dataset that were particularly easy or difficult to detect. In the same way, we wanted to see if there were hard negative candidates that confused a significant portion of the models.

For this, in a first step the detection results of approaches were matched against the ground truth, to yield the list of positives (false negatives and true positives) and false positives. For the matching we used a KDTree-based approach \citep{marzahl2020deep}, where the detection and ground truth centroids were not allowed to exceed an euclidean distance of $7.5\,\mu m$.  

The list of false positives was then, again, grouped using the same proximity criterion using a KD-tree. This was done in order to avoid counting false positives with slightly differing centroid coordinates multiple times, since the false positives can't be assigned to ground truth reference coordinates. We then assigned to each unique false positive (i.e., set of false detections within the distance of $7.5\mu m$) the number of models that opted in favor of this detection. Using this methodology, we were able to calculate the number of votes for each detected cell in the set of detections from all models, and thus judge hard examples from easy ones for the ensemble of all models. 

Additionally, we wanted to evaluate this for the best methods in the field. If we can assume the error to be independent between models, we can hypothesize that the ensemble of the top methods (all performing similarly) could outperform the individual methods. This is also interesting to investigate for ceiling effects in the evaluation caused by label noise in the test set: If the model significantly outperforms the individual models, we can assume that the performance evaluation is currently not limited by such labeling inconsistencies. For this reason, we also constructed an ensemble consisting of all approaches exceeding the baseline, and the baseline. This ensemble of models represents the top five of approaches in the challenge and is denoted as \textbf{top5 ensemble} in the remainder of this work. Since model scores of the individual approaches were unavailable, the ensemble was a simple majority vote. 

\section{Overview of the submitted methods}

All submitted methods used \glspl{CNN} for the task. Table \ref{tab:methods} gives an overview of the network architectures, augmentation, and normalization strategies that were employed. The remainder of this chapter discusses several aspects of the strategies that were employed. 

The detailed descriptions of all methods are published in the proceedings \citep{proceedings}. Here, we want to report on common trends, interesting differences and strategies across the methods to provide readers with an insight on how the task of mitosis detection under domain shift can be tackled. Except for the aspect of translating the task into an (instance) segmentation approach with pixel-level masks, we did not see clear ``winning'' strategies. Instead, we believe that each approach found a strategy that put together matching operators, with some more interchangeable than others. 

\begin{table*}[]
    \centering
    \begin{tabular}{l|l|l|l|l|l}
Team & overall & Scanner A & Scanner D & Scanner E & Scanner F \\ 
\hline
Reference & 0.718 [0.665,0.762]&0.791 [0.673,0.818]&0.708 [0.620,0.766]&0.718 [0.631,0.811]&0.593 [0.551,0.719]\\
CNN baseline & 0.698 [0.639,0.745]&0.687 [0.650,0.812]&0.700 [0.621,0.770]&0.657 [0.519,0.747]&0.606 [0.521,0.717]\\
Top5 ensemble & \textbf{0.773} [0.722,0.813]&0.796 [0.748,0.874]&\textbf{0.745} [0.667,0.780]&0.787 [0.744,0.861]&0.642 [0.581,0.761]\\
\hline
AI medical & \textbf{0.748} [0.704,0.781]&0.793 [0.729,0.830]&\textbf{0.728} [0.643,0.780]&0.781 [0.708,0.843]&0.634 [0.583,0.732]\\
TIA Centre & 0.747 [0.693,0.790]&0.837 [0.692,0.857]&0.677 [0.625,0.759]&\textbf{0.808} [0.683,0.837]&0.667 [0.578,0.768]\\
Tribvn Healthcare & 0.736 [0.670,0.792]&\textbf{0.848} [0.731,0.875]&0.631 [0.608,0.768]&0.795 [0.633,0.848]&0.557 [0.498,0.712]\\
CGV & 0.724 [0.657,0.779]&0.829 [0.750,0.867]&0.643 [0.547,0.728]&0.675 [0.639,0.836]&0.557 [0.521,0.697]\\
XidianU-OUC & 0.707 [0.633,0.768]&0.800 [0.703,0.863]&0.655 [0.528,0.696]&0.795 [0.586,0.814]&0.673 [0.487,0.696]\\
IAMLAB & 0.706 [0.650,0.748]&0.695 [0.646,0.809]&0.721 [0.642,0.757]&0.710 [0.496,0.824]&\textbf{0.690} [0.493,0.681]\\
No.0 & 0.701 [0.637,0.752]&0.826 [0.652,0.837]&0.698 [0.562,0.718]&0.757 [0.571,0.777]&0.632 [0.553,0.696]\\
jdex & 0.696 [0.639,0.739]&0.782 [0.737,0.820]&0.682 [0.575,0.751]&0.667 [0.542,0.741]&0.430 [0.459,0.688]\\
Leeds & 0.686 [0.620,0.737]&0.774 [0.624,0.795]&0.549 [0.503,0.699]&0.696 [0.594,0.786]&0.632 [0.547,0.742]\\
PixelPath-AI & 0.676 [0.615,0.723]&0.620 [0.636,0.788]&0.610 [0.542,0.721]&0.751 [0.607,0.816]&0.576 [0.453,0.667]\\
SK & 0.671 [0.607,0.716]&0.630 [0.636,0.783]&0.644 [0.563,0.730]&0.582 [0.519,0.714]&0.637 [0.464,0.693]\\
ML & 0.632 [0.536,0.713]&0.738 [0.586,0.821]&0.375 [0.267,0.592]&0.755 [0.482,0.820]&0.514 [0.459,0.674]\\
\end{tabular}
    \caption{$F_1$ score for all participating approaches. Numbers in square brackets indicate 95\% confidence interval as determined by bootstrapping. }
    \label{tab:results_table}
\end{table*}
\begin{figure*}
\includegraphics[width=\linewidth]{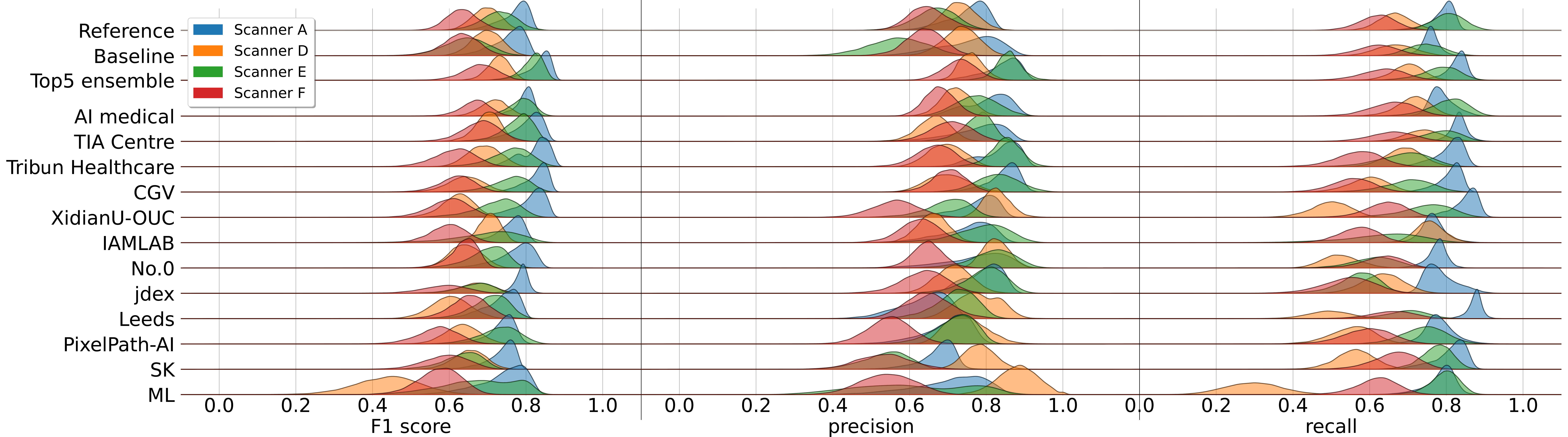}
\caption{$F_1$, precision and recall scores of all participants across scanners (bootstrapping result, 20 random case draws with replacement, 10,000 repetitions).}	
\label{fig:precision_recall_f1}
\end{figure*}

\subsection{Single-stage and multi-stage detection approaches}

Five out of the twelve teams submitted a multi-stage approach for detecting mitotic figures with the first stage generating a list of candidates (targeting a high recall with all mitotic figures included). The second stage then classified whether the extracted patches contained a mitotic figure or not. The first stage was either based on image features~\citep{midog_sk}, or used an object detection network (Faster R-CNN~\citep{midog_pixelpathai}, Mask-R-CNN~\citep{midog_tribun}, U-Net~\citep{jahanifar2021robust}, DetectorRS~\citep{midog_xidianu})). To refine these candidates, ResNet (ResNet50 or larger) or EfficientNet architectures (B0 / B7) were used. Two approaches used Cascade-R-CNN architectures~\citep{midog_no0,midog_iamlab}, which  are inherently multi-stage with sequentially trained detectors and which may therefore be seen as a way to automate the multi-staging. The remaining five teams used a RetinaNet architecture like the reference approach~\citep{midog_cgv,midog_jdex}, the predictions of a U-Net directly~\citep{midog_aimedical,midog_leeds} or a rotation-invariant \gls{CNN}~\citep{midog_ml}.

\subsection{Sampling strategies, loss functions and training mechanisms}
Mitotic figure detection is a highly imbalanced problem and the \glspl{roi} provided in the MIDOG dataset are no exceptions. Most \glspl{roi} contained only 20 or fewer mitotic figures (see \cref{fig:histo_dataset}) and large areas of highly variable  background. Additionally, nuclei, debris or necrotic cells may look extremely similar to mitotic cells (imposters) and therefore pose a considerable challenge whereas other regions can be easily disregarded. This typically prohibits random sampling of the input data and instead requires strategies to deal with this imbalance. Within the submitted approaches, different methods were employed for this, including a focal loss~\citep{lin2017focal} which adjusts the loss for easy samples and was used by~\cite{midog_aimedical,midog_tribun,midog_leeds}, and \cite{midog_jdex}. For segmentation-based approaches, typically a Dice loss or a Jaccard loss was used. Alternatively or additionally, most teams opted for targeted data sampling to ensure sufficient coverage of mitotic figures in each batch, e.g., by randomly undersampling non-mitotic regions~\citep{midog_tiacentre}, excluding regions without mitotic figures~\citep{midog_tribun}, or filtering out easy patches directly~\citep{midog_ml}.

In addition to approaching the task differently (i.e., detection vs. segmentation vs. classification), some teams opted to enrich the task with a domain-adversarial training mechanism ~\citep{midog_no0,midog_sk} similar to the reference approach to encourage domain-independent feature extraction. This was also one of two strategies of using the unlabeled scanner provided in the training set, as alternative to using it for data augmentation via image synthesis.

\subsection{Instance label generation}

The task of the challenge was to find the centroid coordinates of all mitotic figures, which was solved using object detection networks as well as semantic segmentation approaches by most participants. However, three teams \citep{midog_tiacentre,midog_tribun,midog_aimedical} chose to enhance the given set of labels (consisting only of approximate bounding boxes of the mitotic figures) by providing segmentation masks on pixel level for each mitotic figure instance to the training process. 

The approach by \cite{midog_tiacentre} used an CNN-based interactive segmentation model \citep{koohbanani2020nuclick} that is targeted at generating segmentation masks from cell centroid coordinates. The tool is available as an open source tool on GitHub, was trained on publicly available data, and was only used to define segmentation masks for the labels given by the ground truth. 

In a similar way, \cite{midog_aimedical} used another publicly available approach (Hover-Net, \cite{graham2019hover}) that is aimed at nuclei instance segmentation on the dataset. From the output of this tool, they filtered out the segmentation masks of mitoses by thresholding the intersection over union between nuclei detected and ground truth mitotic figure bounding boxes. These segmentation masks were subsequently used as ground truth in a modified U-Net (SK-Unet, \cite{wang2021sk}).

A third strategy was employed in the approach by \cite{midog_tribun}: They generated the segmentation masks for the instances by manually annotating approximately 100 mitoses across the dataset, and then fine-tuned a pre-trained Mask R-CNN on this small dataset to run inference on the remaining annotations and hence derive a segmentation mask for each mitotic figure. The generated instance masks were subsequently used to train another Mask-RCNN model with subsequent secondary classification stage.

\subsection{Data augmentation}
All participants employed some sort of data augmentation or domain adaptation technique during training and/or at test time in order to increase the robustness of their model against the unseen scanner within the final test set.
These techniques where divided in roughly three type of groups:
\begin{itemize}
    \item Standard data augmentation such as: Color, contrast, brightness, geometric
    \item Stain normalization techniques
    \item Generative adversarial networks
\end{itemize}

The standard approach was to apply color, geometric, contrast augmentation during training which should guarantee some level of robustness against unseen data. Groups such as \cite{midog_jdex}; \cite{midog_leeds}; \cite{midog_ml} relied solely on these methods. 
Some approaches used stain normalization techniques either as data augmentation during training or as a way to normalize all of the dataset according to some common stain and then apply data augmentation. Groups such as \cite{midog_tiacentre}; \cite{midog_xidianu}; \cite{midog_iamlab} applied stain normalization to the entire dataset and then proceeded to apply standard data augmentation during training.  

Another approach encountered was using \glspl{gan} to generate images that simulated different scanners and different styles. Groups like \cite{midog_cgv} and \cite{midog_tribun} used a StarGAN and a Residual CycleGAN, respectively, as a data augmentation technique. The benefit of generative adversarial techniques is that they are configurable to simulate a multitude of potential scanner styles, however, the complexity and the hyperparameter search space are increased.

Even though all approaches shared some common characteristics, there was one implementation that stood out in its approach and was, at the end, quite successful in terms of performance: The group \cite{midog_aimedical} used, apart from standard color and geometric transformations of data augmentation, a Fourier-domain transform adaptation approach that separated high-frequent (i.e., structural) from low-frequent (i.e., color) components to transfer the stain information between images, acting as a stain normalization technique without relying on a specific stain matrix transformation.

\subsection{Ensembling and test-time augmentation}
Ensembling combines the outputs of multiple models, either with the same structure or even different model architectures, and is known to improve model robustness as well as overall performance. At the same time, using multiple large parallel models increases the compute time (and also the carbon footprint) at times significantly. While ensembling is common amongst the participants of competitions, we noted much less use of the technique in the MIDOG 2021 challenge. 

Still, five teams employed (moderate) ensembling, mostly using a multi-fold or cross-validation setup on the provided training data and ensembling the resulting models, with different fusion strategies, e.g., simply setting a threshold of necessary detections~\citep{midog_ml}, averaging of classifier predictions~\citep{midog_tiacentre}, or weighted boxes fusion for the detection stage~\citep{midog_pixelpathai}. One team~\citep{midog_tribun} assembled two models for the final classification stage (ResNet50 and DenseNet201) whereas another~\citep{midog_xidianu} put an ensemble of five different models in the center of their classification model. Test time augmentation was less frequently used, but for example by~\cite{midog_tiacentre} for improving classifier performance and by~\cite{midog_tribun} for improving the results of the mask generation during model development.

\begin{figure}
    \centering
    \includegraphics[width=\linewidth]{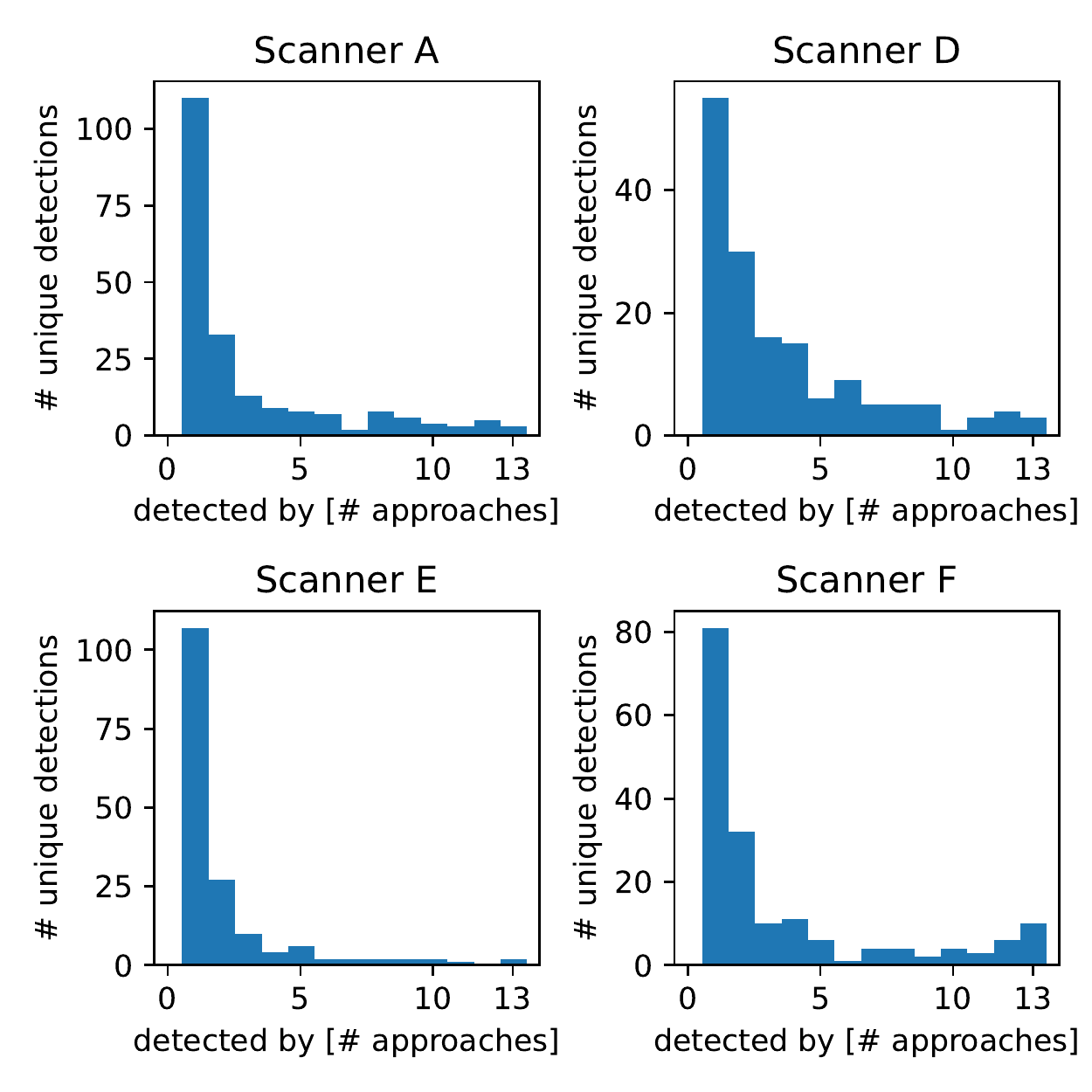}
    \caption{Histogram of false positives across scanners. Note that at least one false detection is necessary for an object to count as false positive. }
    \label{fig:false_positive_histogram}
\end{figure}


\begin{figure}
    \centering
    \includegraphics[width=\linewidth]{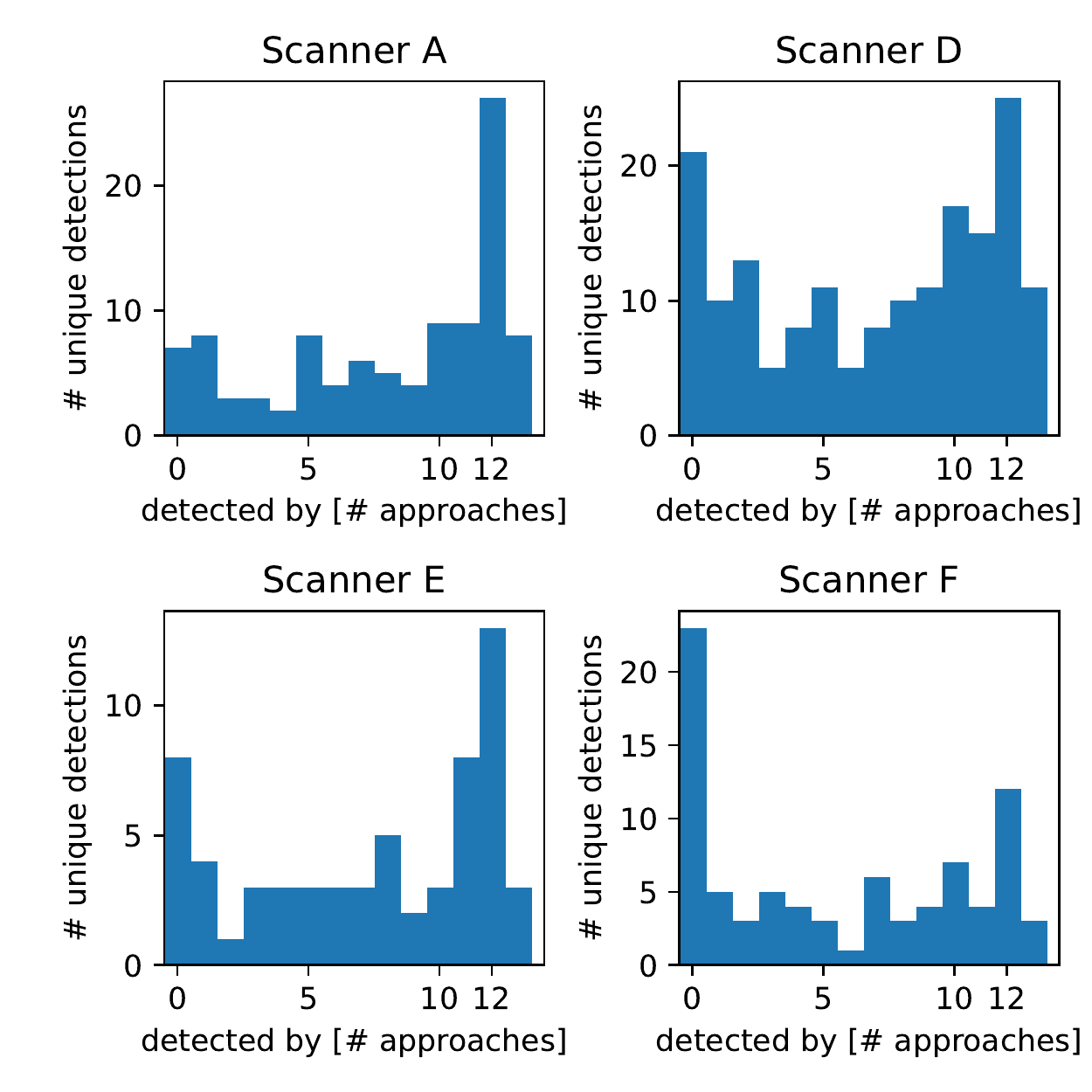}
    \caption{Histogram of detection of ground truth mitotic figures (true positives + false negatives) across scanners, showing how many mitoses have been detected by how many approaches.}
    \label{fig:false_negative_histogram}
\end{figure}

\section{Results}

The majority of approaches were able to provide better results than the CNN baseline. The domain-adversarial reference method \citep{wilm2022} yielded a competitive $F_1$ score of 0.718 on the test set and was outperformed by only four approaches (see Table \ref{tab:results_table}). With an $F_1$ score of 0.748, the overall best performance was reached by Yang \etal, utilizing segmentation and Fourier-domain mixing as augmentation \citep{midog_aimedical}. It is worth noting that this approach was amongst the only approaches which was able to achieve good performance on Scanner D (together with the approach by  \cite{midog_iamlab}), while other approaches performed better on other scanners. Even though the approach did not have the best performance on each individual scanner, it was the most consistently well-performing approach across all scanners, and thus the most generalizing solution. The runner-up approach \citep{midog_tiacentre} had an almost identical $F_1$ score compared to the best approach, supported also by a good overall performance across all scanners, with some minor weaknesses on scanner D, but it was the best performing algorithm on Scanner E. On Scanner F, the best performing approach was the multi-stage cascaded RCNN approach by~\cite{midog_iamlab}, and on the scanner A, which was part of the training set, the best solution came from \cite{midog_tribun}. 

\subsection{Post-Challenge Ensembling}

The top5 ensemble outperformed the leading approaches by \cite{midog_aimedical} and \cite{midog_tiacentre} in terms of overall $F_1$ score considerably. As Table \ref{tab:results_table} shows, this can be mainly attributed to a boost in performance for Scanner D, where the margin to the runner-up approach is the largest. For the other scanners, the top5 ensemble is on par with the respectively best approach for the scanner (especially when the 95\% confidence intervals are considered). This ensembling method also yielded the overall highest precision (see \cref{fig:precision_recall_f1}) on all scanners.   

\subsection{Object-level agreement}

\begin{figure}
    \centering
    \includegraphics[width=\linewidth]{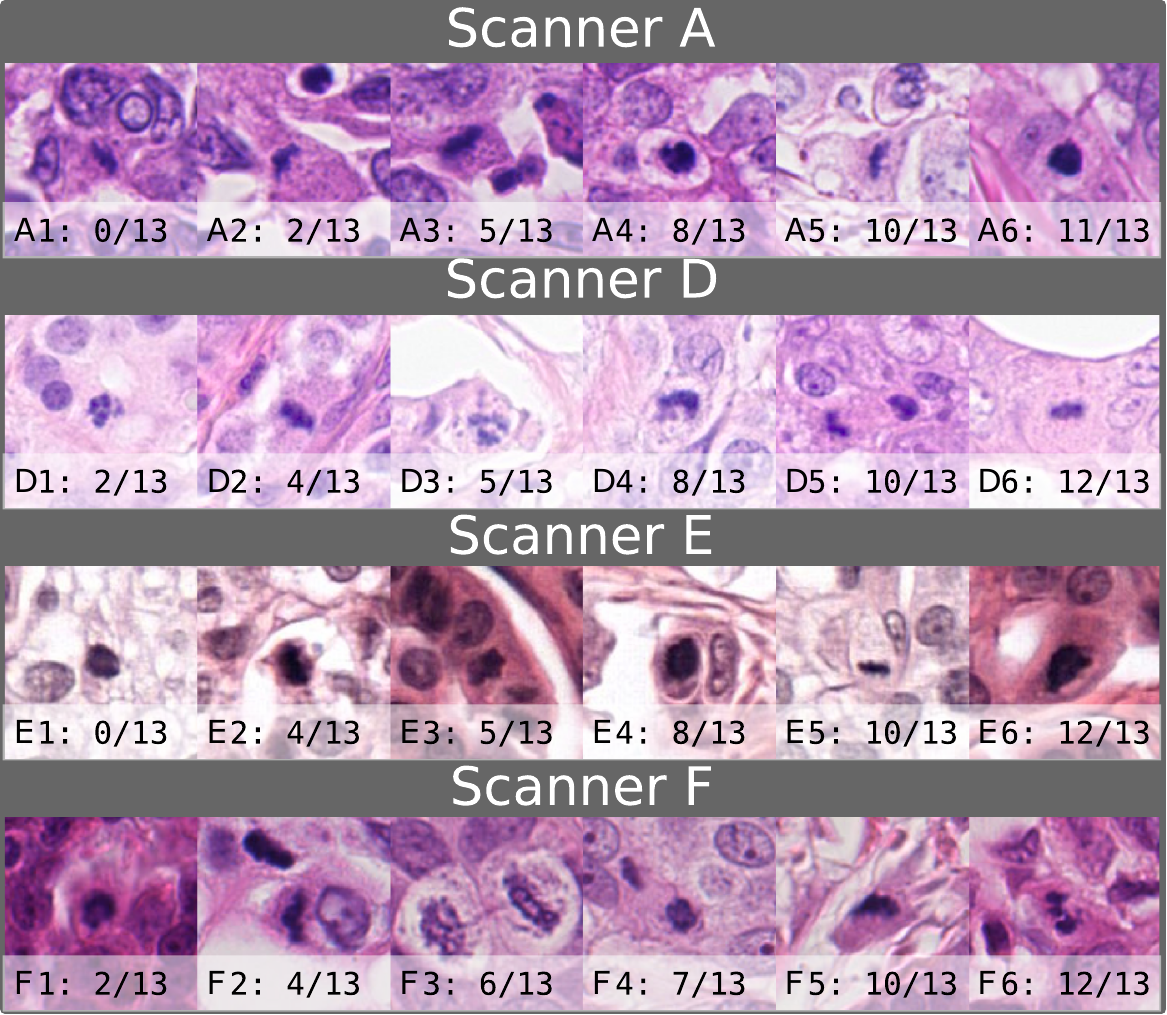}
    \caption{Examples of ground truth mitotic figures (true positives and false negatives), ordered by the count of models voting for it. The numbers (x/13) indicate, how many models voted for this cell to be a mitotic figure. The rows are stratified by the number of models to give examples for the complete distribution in \cref{fig:false_negative_histogram}.}
    \label{fig:false_negative_examples}
\end{figure}

\begin{figure}
    \centering
    \includegraphics[width=\linewidth]{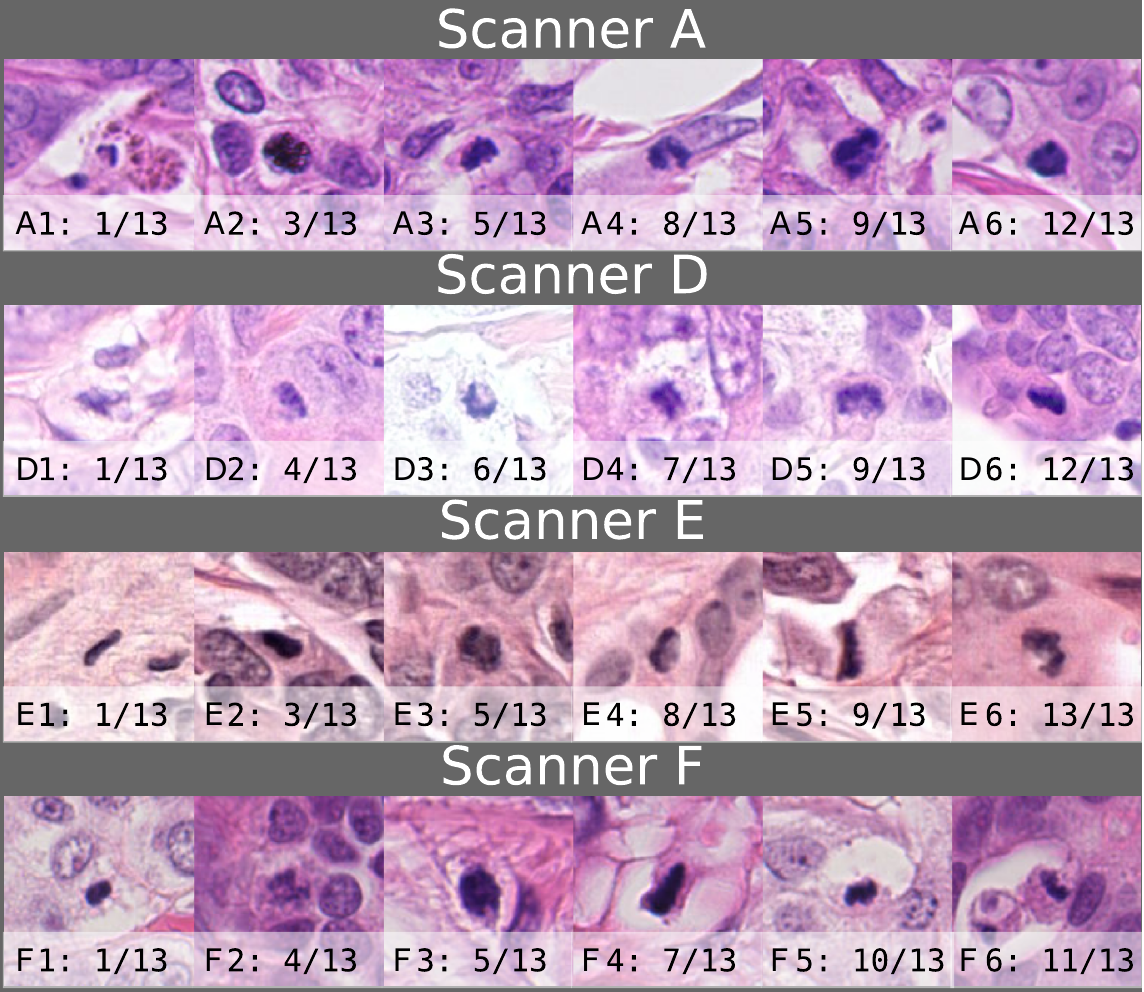}
    \caption{Examples of false positives, ordered by the count of models voting for it. The numbers (x/13) indicate, how many models voted for this cell to be a true mitotic figure.}
    \label{fig:false_positive_examples}
\end{figure}

To investigate the diversity in detections on object level, we assessed the agreement between the approaches on false positives (non-mitotic figure objects found by one or multiple approaches) and on the ground truth mitotic figures. The histogram of false positives is given in \cref{fig:false_positive_histogram}. It shows that the vast majority of unique false detections was only detected by a small number of approaches. There is little difference in this behavior across scanners. Looking at the histogram of false negatives in \cref{fig:false_negative_histogram}, we see a different behavior for missed detections: While most out of the total set of mitoses were only missed by a small number of methods for the seen scanner A (as visible in the high counts for mitoses that were found by $>$10 models), the number of mitotic figures that were missed by the majority or even the totality of detectors increased for the unlabeled/unseen scanners D and F, and also for scanner E (note that this scanner had a lower overall MC as of the ground truth labels). This is also underlined by the generally lower recall for those scanners compared to the seen scanner A (see \cref{fig:precision_recall_f1} and Table \ref{tab:results_table}). The total number of mitoses detected by all of the approaches was 25, i.e. the vast majority of mitoses was missed by at least one approach. In \cref{fig:false_negative_examples} and \cref{fig:false_positive_examples}, we give examples for false positives and ground truth mitosis (true positives and false negatives), stratified by the number of models that detected those. It becomes obvious that some of those might be borderline mitotic figures and might even have a different label when re-evaluated by the same or a different set of experts (e.g., examples A1, D1, E3, or even F5 in \cref{fig:false_negative_examples}). On the other hand, the low contrast of scanner D seemed to be a major obstacle for many approaches, even though sample images (without labels) of the scanner were available within the training set. Also atypical visual representations of clear mitoses, such as the examples D3 (atypical) or F3 (late telophase) were apparently hard to detect, as well as cells with unclear cell boundaries (A1, A2, D2). 

Looking at some of the examples of false positives in  \cref{fig:false_positive_examples}, we see that especially the cells detected by most models (e.g., A3, A4, A6, D6, E6, F6) can be considered missed annotations of the dataset. Cellular objects incorporating bar-like structures, such as in the examples E5, F4) can be considered hard negatives for most of the approaches.

As depicted in \cref{fig:precision_recall_f1}, the top performing models excelled at finding a good compromise between precision and recall across all scanners. The unknown color distribution of the unseen scanners can be particularly expected to have an influence on the calibration of the model output and thus result in underestimated or overestimated model scores, an effect which can be observed especially for the results on scanner D for the approaches by \cite{midog_xidianu}, \cite{midog_iamlab}, \cite{midog_no0}, \cite{midog_sk}, and \cite{midog_ml}. 

It is worth noting that all three best performing approaches included an auxiliary task for mitosis segmentation, which apparently increased general performance. From the results, it can, however, not be determined if the auxiliary task also helped in domain generalization.  Five out of the seven top performing approaches were utilizing methods of ensembling or test-time augmentation. Further, we see a slight trend that the better performing methods had larger classification networks.

\section{Discussion}

The results of the MIDOG 2021 challenge indicate that, using proper augmentation strategies and deep learning architectures, domain shift between whole slide imaging scanners can be compensated for to a high degree. The results by the best approaches in the field were in the range of well-performing human experts on the same task (see \cite{Bertram2021VetPathol}). It must be noted, however, that the mitosis detection task was only performed on selected \glspl{roi}. In contrast, \glspl{wsi} will have a much higher variability in tissue quality, including out-of-focus areas, and areas with necrosis, and thus have a variety of hard negative examples for the algorithms that have not been evaluated in this challenge. 

All three expert pathologists were highly familiar with mitotic figure identification, however, a bias in annotation can not be excluded completely. Ultimately, a prediction of outcome, such as survival or recurrence, based on mitotic figure detection on \glspl{wsi}, complemented with other morphological factors, would be the clinical target for an automated tumor grading. Yet, since this was not the scope of the MIDOG challenge, this evaluation is considered future work.

The best achieved $F_1$ score on Scanners A (fully supervised) and E (unknown) was in the order of state-of-the-art approaches trained fully supervised in-domain \cite{aubreville2020completely,bertram2019large}. In contrast, the best results on Scanner D (image only, no labels provided) and Scanner F (unknown) were considerably weaker. While the good performance on Scanner A underlines the high consistency of labels due to the computer-aided approach and high level of expertise of our pathologists, the weak performance on scanners D and F might be related to the domain shift not being covered completely by the algorithms. On the other hand, it might be influenced by the image quality of the scanners (which might make mitosis detection in general more challenging for humans and algorithms alike) or even non-familiarity of the experts with the color and structural patterns representing mitoses within tissue imaged by those scanners. Thus, while this challenge evaluated mitosis detection on the largest set of scanners with controlled staining conditions, the evaluation on this subset of scanners might still have its limitations and not generalize to other, yet unseen scanners. 

Even though most of the higher ranked approaches used ensembling or test-time augmentation, it is unclear if those methods were a success factor in our challenge setup or if there is a mere interrelation between participants utilizing ensembling / TTA and a higher algorithmic development effort, reflected in a higher score. However, the superior performance of the top5 ensemble (as an ensemble of approaches) points into the direction that, in general, ensembling techniques are a success factor for these kind of tasks.

The MIDOG challenge was the first in the field of generalization of mitosis detection to unseen domains, and thus an important step towards clinical applications. And yet, we can observe, that there are many challenges ahead on route to a clinical application: While, as mentioned, application on \glspl{wsi} is a very different challenge, required for clinical application, so is the generalization to further tissue and cancer types, where mitosis detection plays an equally important role in the respective grading system. In fact, we can expect a substantial domain shift between tumor types, which is why this task will be the focus of the successor event of this challenge.

\bibliographystyle{model2-names.bst}\biboptions{authoryear}
\bibliography{bibliography}

\section*{Acknowledgement}
Computational resources for the challenge have been donated by NVIDIA and AWS. The authors would like to thank the team from grand-challenge.org for their expertise and support. The challenge organizers would like to thank Siemens Healthineers for donating the monetary prizes of the challenge, who have been awarded to the top three participants.

\section*{Conflicts of interest}
The authors declare no conflicts of interest. 

\section*{Data usage statement}
All data of the training set was released under the Creative Commons 4.0 BY (attribution) NC-ND (non-commercial, non-derivative) license.

\appendix

\section{Author contributions}
The challenge was organized by Katharina Breininger, Natalie ter Hoeve, Christof A. Bertram, Francesco Ciompi, Robert Klopfleisch, Andreas Maier, Nikolas Stathonikos, Mitko Veta, and Marc Aubreville.

Frauke Wilm, Christian Marzahl, Katharina Breining and Marc Aubreville provided the algorithmic reference approach for the challenge.

The core writing group of this paper consisted of Marc Aubreville, Nikolas Stathonikos, Christof A. Bertram, Katharina Breininger, and Mitko Veta. 

Taryn A. Donovan, Robert Klopfleisch and Christof A. Bertram served as expert pathologists in annotating the complete challenge data set. 

Jack Breen and Nishant Ravikumar (Team Leeds), Youjin Chung and Jinah Park (Team CGV), Ramin Nateghi and Fattaneh Pourakpour (Team PixelPath-AI), Rutger H.J. Fick and Saima Ben Hadj (Team Tribun Healthcare), Mostafa Jahanifar and Nasir Rajpoot (Team PixelPath-AI), Jakob Dexl and Thomas Wittenberg (Team jdex), Satoshi Kondo (Team SK), Maxime W. Lafarge and Viktor H. Koelzer (Team ML), Jingtang Liang and Yubo Wang (Team XidianUOUC), Xi Long and Jingxin Liu (Team No. 0), Salar Razavi and April Khademi (Team IAMLAB), and Sen Yang and Xiyue Wang were the respective first and last authors of the participant’s challenge papers, and thus contributed the algorithmic approaches to the challenge. 

All authors reviewed the manuscript.

\end{document}